 \documentclass[pmlr,twocolumn,10pt]{jmlr} % W&CP article

 \usepackage{placeins}

\usepackage{booktabs}
\usepackage{siunitx}

\usepackage[switch]{lineno}

\theorembodyfont{\upshape}
\theoremheaderfont{\scshape}
\theorempostheader{:}
\theoremsep{\newline}

\jmlrvolume{259}
\jmlryear{2024}
\jmlrsubmitted{LEAVE UNSET}
\jmlrpublished{LEAVE UNSET}
\jmlrworkshop{Machine Learning for Health (ML4H) 2024} % W&CP title

\title{Investigating Primary Care Indications to Improve Electronic Health Record in Dementia Target Trial Emulation}

\author{%
\Name{Max Sunog} \Email{msunog@mgh.harvard.edu}\\
\addr Massachusetts General Hospital, US
\AND
\Name{Colin Magdamo} \Email{colin\_magdamo@hms.harvard.edu}\\
\addr Harvard Medical School, US
\AND
\Name{Marie-Laure Charpignon} \Email{mcharpig@mit.edu}\\
\addr MIT Institute for Data, Systems, and Society \& Schmidt Center at the Broad Institute of MIT-Harvard, US
\AND
\Name{Mark Albers} \Email{malbers@mgh.harvard.edu}\\
\addr Massachusetts General Hospital \& Harvard Medical School, US
}

\begin{document}

\maketitle

\begin{abstract}
Missing data, inaccuracies in medication lists, and recording delays in electronic health records (EHR) are major limitations for target trial emulation (TTE), the process by which EHR data are used to retrospectively emulate a randomized control trial. EHR TTE relies on recorded data that proxy true drug exposures and outcomes. We investigate the under-utilized criterion that a patient has indications of primary care provider (PCP) encounters within the EHR. Such patients tend to have more records overall and a greater proportion of the types of encounters that materialize comprehensive and up-to-date records. We examine the impact of including a PCP feature in the TTE model or as an eligibility criterion for cohort selection, contrasted with ignoring it altogether. To that end, we compare the estimated effects of two first line antidiabetic drug classes on the onset of Alzheimer’s Disease and Related Dementias (ADRD). We find that the estimated treatment effect is sensitive to the consideration of a PCP feature, particularly when used as an eligibility criterion.  Our work suggests that this PCP feature should be further researched.
\end{abstract}

\paragraph*{Data and Code Availability}
The study uses EHR data from the Research Patient Data Registry \citep{rpdr-07}, social vulnerability index (SVI) data from the Agency for Toxic Substances and Disease Registry \citep{svi}, and Massachusetts death records from the Registry of Vital Records and Statistics. Because the data contain patient information, they cannot be made available. The code is available in the supplement.

\paragraph*{Institutional Review Board (IRB)}
This research was performed under MGB IRB approval (protocol 2023P000604). 

\section{Introduction}
\label{sec:intro}

The widespread use of electronic health records (EHR) for collecting healthcare information has generated large stores of data. Through the target trial emulation (TTE) framework, these data enable otherwise infeasible studies when traditional randomized control trials are prohibited due to ethics, recruitment difficulties, or long trial duration. For example, testing drug repurposing hypotheses to delay ADRD onset is greatly facilitated by EHR TTE; one such study found a protective effect on ADRD onset for patients recorded as initiating the antidiabetic metformin vs. sulfonylureas \citep{charpignon-22}.

In EHR TTE, patients are \emph{enrolled} based on records of treatment initiation, and their follow-ups and censorship dates are \emph{derived} from records of an outcome or last visit. Because these are indirect observations, the estimated risk is that of an \emph{outcome being recorded}, given the \emph{recorded treatment(s)}. The assumption is that such an estimated risk reflects the risk that relates the true treatment assignment to the true outcome; for instance, the above result that patients with an initial metformin prescription record have a lower risk of recorded ADRD onset relative to sulfonylureas suggests that patients who truly initiate metformin similarly have a lower risk of developing ADRD, relative to true sulfonylurea initiators.

Under this premise, methodological work has strengthened the relationship between recorded treatment and recorded outcome. For example, considering competing risks (e.g. death before the main outcome) has reduced bias of risk estimates for the primary recorded outcome \citep{anderson-93}.

However, these methods do not address the assumption that EHR data are reflective of events that occur in nature. Therefore, accurate and timely records are needed to interpret EHR TTE findings as real-world effects that can inform clinical actions, like switching a patient’s treatment to metformin because of a protective effect against ADRD. Unfortunately, missed or unrecorded diagnoses, inaccurate drug lists, and severe delays in diagnosis recording are common. As a result, EHR phenotyping often lacks sensitivity: a meta-review of algorithms using structured EHR data to phenotype dementia found that sensitivities ranged from 8\% to 79\% compared to expert clinical evaluation or chart review \citep{walling-23}. While broadly problematic, this missingness can also bias results because patients in minority groups or with lower socioeconomic status are more likely to have missing records \citep{getzen-23}.

To address this limitation, other approaches must be employed to improve data quality. Principal factors associated with higher quality EHR data are 1) many recorded encounters, 2) types of procedures that inform thorough medical records (taking patient history, running regular screens, etc.), and 3) interactions with a provider who actively inputs records into the EHR \citep{verheij-18}. Given these conditions, one under-utilized metric is whether a patient sees a Primary Care Physician (PCP) within the EHR’s healthcare system. At annual wellness visits, PCPs are likely to document a complete review and comprehensive history of their patients, and to update their EHR \citep{sleath-99}. While prior work has utilized the total level of healthcare utilization in EHR TTE \citep{goldstein-16}, this study adds a complementary method that critically accounts for the types of encounters.

In this study, we 1) developed a refined definition of internal PCP utilization by mining the EHR, 2) demonstrated improved data recording quality among patients that meet the PCP definition, and 3) explored the effects on TTE results of using these indications in modeling and eligibility criteria.  These changes in the estimated treatment effect stem from reduced bias because of the more thorough history and follow-up for patients with PCP encounters.

\section{Methods}

\subsection{Defining the PCP indication}

We identified three useful types of PCP indications present in our EHR system, the 
Research Patient Data Registry (RPDR): 1) procedure codes associated with primary care visits (e.g., annual wellness exams); 2) encounters under the ‘Primary Care’ service line, a categorization for visits performed by a PCP; and 3) encounters with ‘Annual Wellness Visit’ as the listed reason-for-visit (code attached to encounters logged in Epic). Using these indications, we defined a composite metric: a patient was considered to have an internal PCP if they had at least one such indication before treatment initiation.

In our cohort of patients with type 2 diabetes (T2D), individuals who met the internal PCP definition had higher healthcare utilization and recorded disease prevalence rates than those who did not (see Appendix B). When comparing the comorbidity distribution of our cohort with that of adults with T2D in prior observational studies using well phenotyped medical histories, we found a better alignment between the PCP group and the literature, while the no-PCP group had consistently lower prevalence rates.
For instance, while 10\%-20\% of patients with T2D were found to also suffer from COPD \citep{mamillapalli-19}, the PCP and no-PCP groups had prevalence rates of 6.0\% and 3.0\%, respectively. Similarly, 49.1\% of T2D patients are obese \citep{nguyen-10}; our PCP and no-PCP groups have obesity prevalence rates of 56\% and 21\%. Although the populations captured in prior studies do not exactly match our cohort, the better alignment with the PCP group suggests that patients who have an internal PCP have a more complete EHR than those who have not.

Notably, PCP patients had half the death rate of those without a PCP. To investigate this, we compared EHR death records with the MA state death registry, an accurate source for records of deaths that occur in MA. We use this source in TTE to offset missingness in our EHR, which is expected for some recent deaths.
Among the roughly 7,000 patients in our cohort with a record in the MA death registry, 46.3\% of no-PCP patients and 28.6\% of PCP patients are missing corresponding death records in the EHR, so PCP patients have less missingness in their EHR death records. Therefore, the higher recorded death rate for no-PCP patients implies that their true death rate is elevated. In fact, the age-specific death rate for PCP patients aligned closely with those in the public life tables for the entire MA population \citep{arias-22}, while the no-PCP patients consistently had roughly twice the death rate (see appendix I). 

Finally, we compared our PCP and no-PCP cohorts to a study reporting age-specific ADRD incidence per person-year using Medicare claims data for over 8 million patients \citep{olfson-21}. Although our population differs in racial demographics and by being restricted to T2D patients, we expect that if we had perfect outcome recording, a replication of their methodology would result in similar incidence rates. For ages up to 75, both of our cohorts aligned well with the study. Above 75 – when ADRD is most common and alignment most critical – the PCP cohort had similar incidence rates, while the no-PCP cohort consistently had roughly 60\% the reported rates (see appendix H), suggesting that ADRD diagnosis missingness is far more common for the no-PCP patients.

\subsection{Methods of utilizing the PCP indication}

To investigate the sensitivity of TTE results to the approach used to handle the PCP indication, we replicated the prior metformin vs. sulfonylureas emulation. In the original study, 13,191 participants from the Research Patient Data Registry (RPDR) with incident antidiabetic prescriptions between January 2007 and September 2018 were selected. In our replication, we used the same indication code lists, eligibility criteria, and covariates, all of which are listed in appendix C. From this starting point, we tested three sensitivity analyses comparing methods of using the PCP indication to increase data quality: a baseline strategy ignoring PCP indications (B), a modeling strategy adding a PCP covariate in the propensity model (M), and an exclusion strategy requiring a PCP indication for inclusion in the cohort (E). Aside from using the PCP indication, our study diverged by including data through April 2024. (B) and (M) had a cohort of N=54,440 (17,118 PCP patients and 37,322 no-PCP patients) and (E) had N=17,118, both larger than the original study’s.

The PCP strategy was evaluated across these three methodologies by comparing the hazard ratio (HR) obtained from a Cox proportional hazards model and cumulative incidence functions (CIF) that account for the competing risk of death \citep{getzen-23}.

\section{Results}
\label{sec:math}

\begin{figure}[]
 % Caption and label go in the first argument and the figure contents
 % go in the second argument
\floatconts
 {fig:nodes}
 {\caption{\small Hazard ratios and $95\%$ confidence intervals for the treatment effect on ADRD incidence, estimated from Cox models.}}
 {\includegraphics[width=0.9\linewidth]{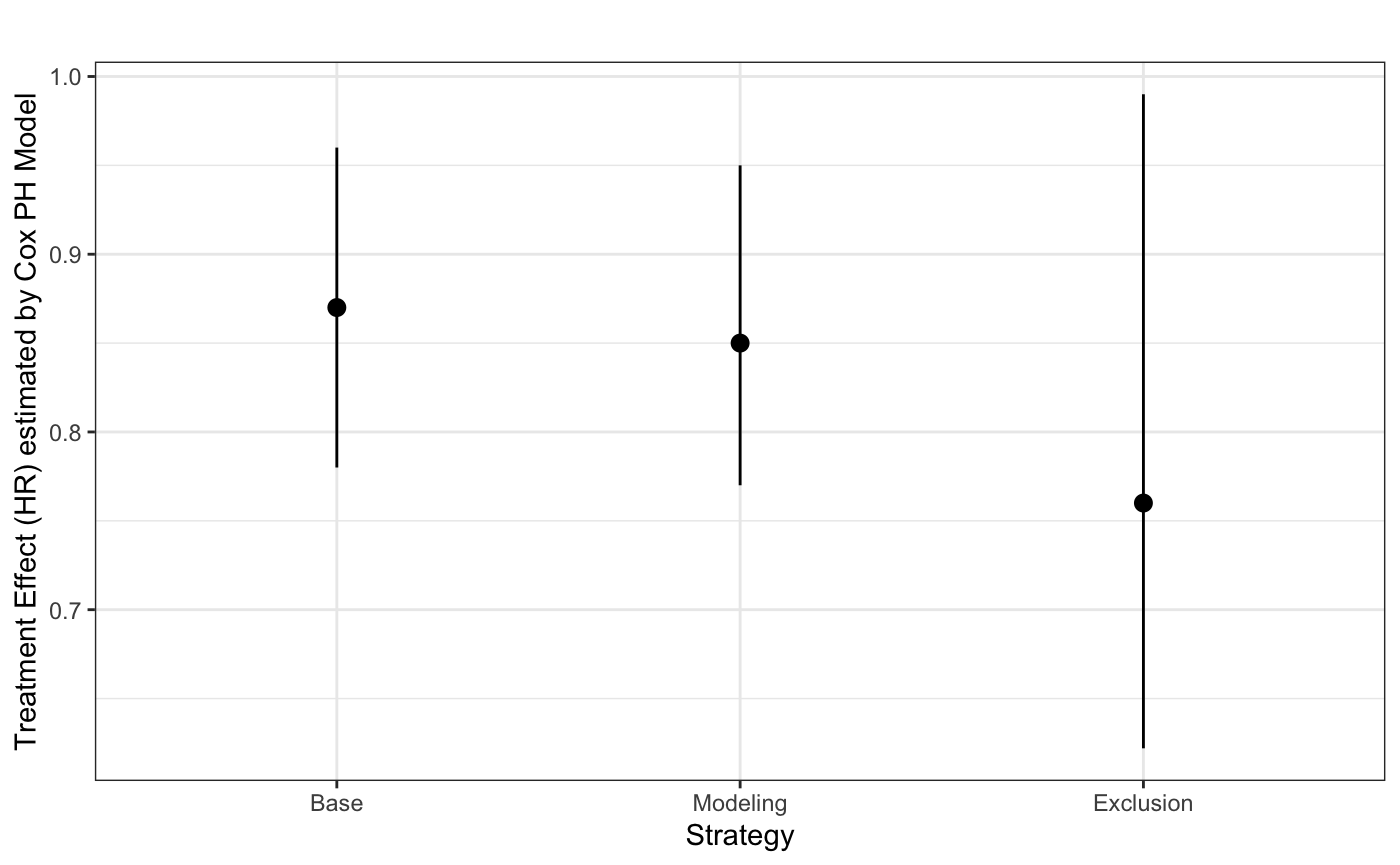}}
\end{figure}

We found that the HR estimates for the effect of initiating metformin were similar in (B) and (M): .87 (95\% CI: .78 - .96, p=.006) and .85 (95\% CI: .77 - .95, p=.003), respectively. In (E), the HR estimate was .76 (95\% CI: .61 - .95, p=.016), an expected increase in confidence interval width, given the much smaller cohort size after selecting on the PCP indication. The stronger effect in (E), which is outside the confidence intervals of both other experiments, may be due to reducing confounding by eliminating patients with less complete and accurate EHR data.

\begin{figure}[htbp]
 % Caption and label go in the first argument and the figure contents
 % go in the second argument
\floatconts
 {fig:nodes}
 {\caption{\small Cumulative incidence functions for ADRD (top) and death before ADRD (bottom) for each strategy, comparing the sulfonylurea (red) and metformin (blue) arms. }}
 {\includegraphics[width=.9\linewidth]{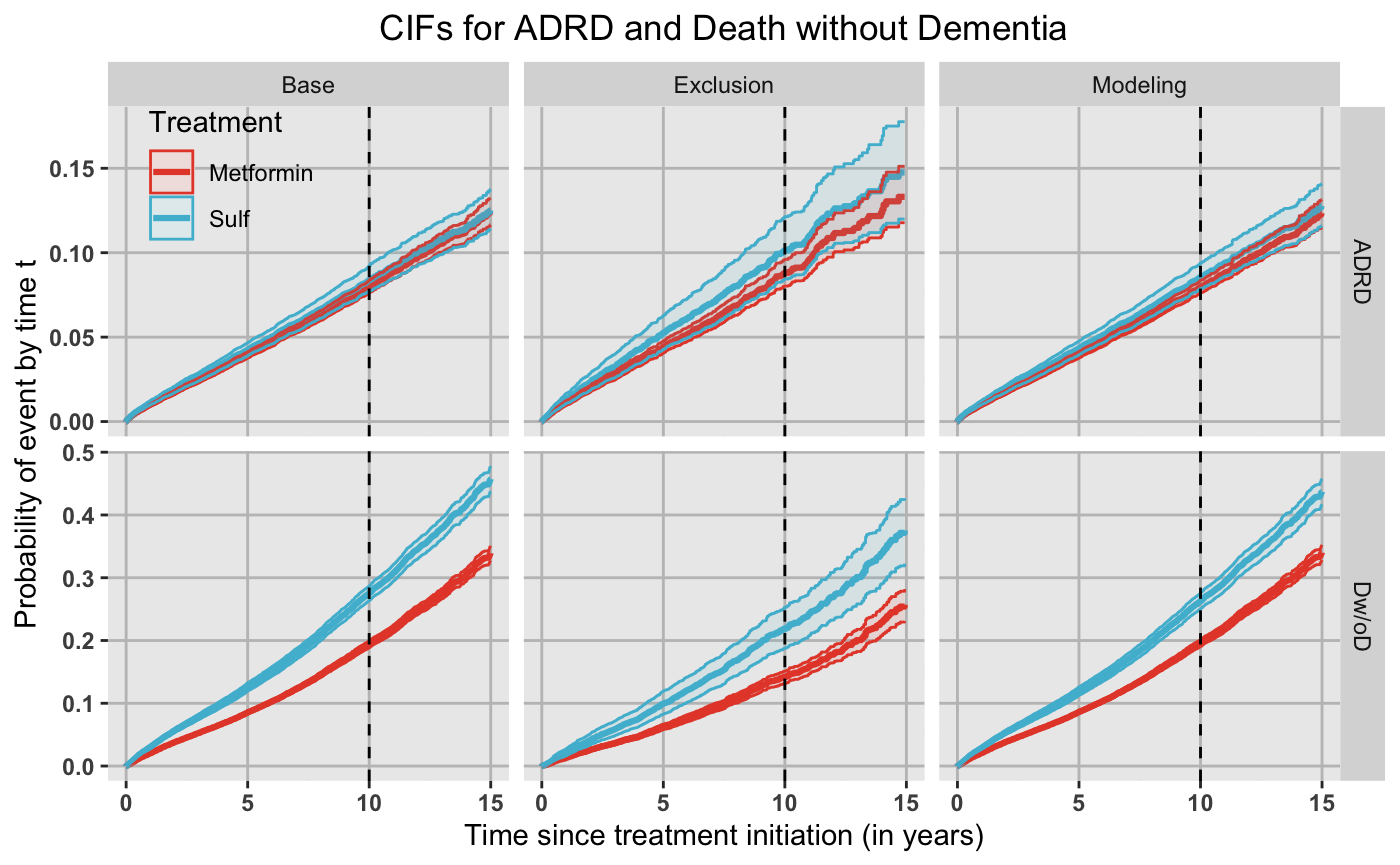}}
\end{figure}

To quantify the treatment effect accounting for the competing risk of death, the risk difference across treatment arms at ten years after baseline ($RD_{10}$) was examined. The death-without-dementia (D{\O}D) curves for the two arms are slightly closer in (M) than (B): the $RD_{10}$ of D{\O}D were -8.1\% (95\% CI: -9.5\% - -7.0\%) and -6.7\% (95\% CI: -8.0\% - -5.4\%), respectively. As the competing risk outcome model does not use covariates, this effect is from the propensity model. The distribution of PCP indications is disproportionately higher in metformin patients, so this suggests that decreasing the weight of metformin patients with a PCP indication attenuates the mortality survival-time effect. For the ADRD curves, the $RD_{10}$ are -.39\% (95\% CI: -.91\% - .19\%) in (B) and -.58\% (95\% CI: -1.4\% - .17\%) in (M), attributing a minor increase in the effect of metformin.

In (E), the ADRD curves for both treatments increase, which is far more likely to be the result of higher outcome ascertainment in the entire cohort than of a higher rate of actual cognitive decline among patients with a PCP indication. As observed in the Cox model, the smaller sample size results in much larger confidence intervals. In this experiment, the $RD_{10}$ of D{\O}D is -7.7\% (95\% CI: -11\% - -4.5\%), between that of the other experiments. Notably, the $RD_{10}$ of ADRD is -1.3\% (95\% CI: -3.2\% - .55\%), more than three times that found in (B).

\section{Discussion}

To formulate our definition of internal PCP encounter, we chose indications based on a manual review of the available codes. Other US EHRs may require a different definition; we are exploring this in a California EHR. We found that the PCP indication significantly impacted the results of an EHR TTE, especially when used as an eligibility criterion. Here, this exclusion strategy is desirable because the ADRD CIFs are less noisy when lower quality data are removed instead of balanced between arms. Other TTEs may prefer the modeling strategy, e.g., if applying stricter eligibility criteria excludes too many patients.

Notably, our PCP criterion is not the only utilization-based method to improve data quality and it should preferably be used in conjunction with other methods. For instance, \citet{goldstein-16} demonstrated with simulated and EHR data that outcome adjustment on a patient’s number of encounters in the EHR meaningfully changes the odds ratio between the recordings of two conditions. However, the authors note that one cause of incomplete recording is patients moving between healthcare systems, which motivates more work to identify ways of ensuring data quality. Combining adjustment for healthcare utilization with the PCP criterion can be beneficial, particularly as the PCP criterion accounts for visit types.

However, employing healthcare utilization features requires care, as they can act as colliders. Utilization is often affected by treatments and outcomes, so conditioning on such features can induce a selection bias \citep{weiskopf-23}. Fortunately, the PCP feature should result in minimal bias, as primary care encounters are less correlated to specific medical conditions or events \citep{weiskopf-23}.

When using the PCP criterion, there is potential for bias because patients with a PCP have higher educational attainment and socioeconomic status overall \citep{getzen-23}. However, nearly all patients in the US who receive a drug prescription for a chronic disease – such as metformin or sulfonylureas for T2D -- will have a PCP, so members of a drug repurposing TTE excluded by the internal PCP criterion likely receive primary care outside of the EHR’s network. In our cohort, patients with and without an internal PCP had similar educational attainment, suggesting that the selection bias associated with considering only patients with an internal PCP may be limited. Additionally, the findings of this TTE are clinically relevant for patients eligible to receive an antidiabetic prescription and reachable, so a cohort restricted to patients with a PCP may better represent the population of interest.

While EHR data are a valuable source of information, EHR studies are inherently unrepresentative of the general population. A key step in working towards generalizability is reporting demographic data more comprehensively \citep{boyd-23}, which can be worked towards with the PCP feature. Because EHR data are disproportionately missing for less privileged groups, the distribution of patients who appear at all in the EHR is not reflective of those with thorough recording, so ignoring the PCP criterion may obscure the latent bias. Future work to improve data quality should further refine the PCP criterion, e.g., by incorporating unstructured provider notes in the EHR. To mitigate biases in the EHR and fully benefit from the PCP criterion, studies should expand generalizability through replications in other EHRs, and more broadly with public efforts to promote healthcare accessibility for marginalized groups.

\section{Citations and Bibliography}
\label{sec:cite}

\acks{The authors thank other members of the DRIAD-EHR team including Bella Vakulenko-Lagun, Deborah Blacker, Sudeshna Das, Jeff Klann and Shawn Murphy for thoughtful comments. This study was funded by the NIH R01 AG058063 (awarded to M.W.A.).}

\bibliography{manuscript}

\newpage

\appendix

\onecolumn
\section{Venn diagram visualizing the number of patients in the full cohort with each type of internal PCP utilization indication prior to baseline.}\label{apd:first}

Percentages listed under the names are proportions of the full cohort; percentages within the circles are proportions of all PCP patients.
\begin{figure*}[htbp]
\floatconts
 {fig:nodes}
 {\caption{Venn Diagram of Patients with each PCP Indication}}
 {\includegraphics[width=1\linewidth]{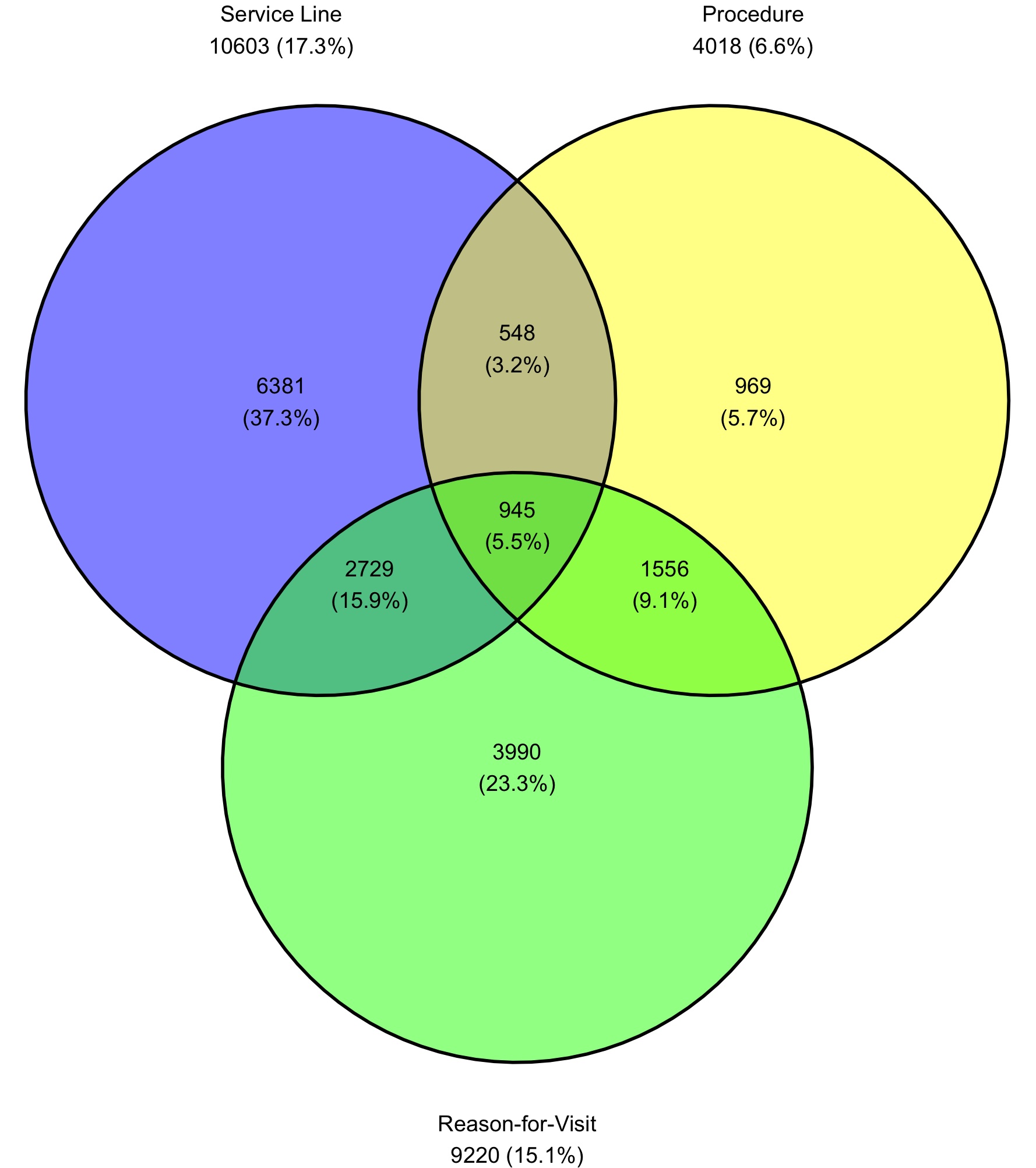}}
\end{figure*}

\FloatBarrier

\newpage
\section{Expanded Table of Summary Stats}

\subsection{Table by PCP Status}

\begin{table*}[htbp]
\floatconts
 {tab:example}%
 {\caption{Full comparison of baseline summary statistics, stratified by whether a patient has a record of an internal primary care encounter prior to baseline.}}%
 {\begin{tabular}{lll}%
 \bfseries Feature & \bfseries No PCP & \bfseries PCP\\ 
 Total (N) & $37,322$ & $17,118$ \\ 
 AD & $6.0{\%}$ & $5.0{\%}$ \\ 
 Death & $21.0{\%}$ & $10.0{\%}$ \\ 
 Age (mean) & $66.520$ (sd: 9.6) & $64.480$ (sd: 8.9) \\ 
 Sex Female & $50.0{\%}$ & $53.0{\%}$ \\ 
 Education Secondary & $32{\%}$ & $37{\%}$ \\ 
 Education College & $35{\%}$ & $39{\%}$ \\ 
 Education Graduate & $08{\%}$ & $9{\%}$ \\ 
 Education Missing & $25{\%}$ & $16{\%}$ \\ 
 Socioeconomic Vulnerability Score (0-1) & $0.32$ (sd: .21) & $0.36$ (sd: .24) \\ 
 Home Life Vulnerability Score (0-1) & $0.44$ (sd: .20) & $0.47$ (sd: .21) \\ 
 Racial Ethnic Vulnerability Score (0-1) & $0.41$ (sd: .21) & $0.48$ (sd: .23) \\ 
 Housing Vulnerability Score (0-1) & $0.50$ (sd: .18) & $0.54$ (sd: .18) \\ 
 Hypertension Diagnostic Code & $56{\%}$ & $81{\%}$ \\ 
 Stroke Diagnostic Code & $4{\%}$ & $6{\%}$ \\ 
 Cancer Diagnostic Code & $30{\%}$ & $39{\%}$ \\ 
 COPD Diagnostic Code & $3.0{\%}$ & $6.0{\%}$ \\ 
 Overweight Diagnostic Code & $4.0{\%}$ & $22{\%}$ \\ 
 Obesity Diagnostic Code & $21.0{\%}$ & $56.0{\%}$ \\ 
 Cardiovascular Disease Diagnostic Code & $17.0{\%}$ & $26.0{\%}$ \\ 
 Visits Before Baseline (mean) & $34$ (sd: 50) & $128.200$ (sd: 123) \\ 
 Visits Year Before Baseline (mean) & $6.230$ (sd: 11) & $17.250$ (sd: 17) \\ 
 Outpatient Visits Before Baseline (mean) & $27.820$ (sd: 44) & $116.180$ (sd: 115) \\ 
 Outpatient Visits Year Before Baseline (mean) & $5.270$ (sd: 10) & $16.500$ (sd: 17) \\ 
 \end{tabular}}
\end{table*}

\FloatBarrier

\subsection{Table by Treatment Arm}

\begin{table*}[htbp]
\floatconts
 {tab:example}%
 {\caption{Full comparison of baseline summary statistics, stratified by treatment arm. Includes both PCP and no-PCP patients.}}%
 {\begin{tabular}{lll}%
 \bfseries Feature & \bfseries Metformin & \bfseries Sulfonylurea\\ 
 Total (N) & $46,613$ & $7,826$ \\ 
 PCP (N) & $15,910 (34.1\%)$ & $1,208 (15.4\%)$ \\ 
 AD & $4.9{\%}$ & $8.8{\%}$ \\ 
 Death & $14.0{\%}$ & $36.6{\%}$ \\ 
 Age (mean) & $65.1$ (sd: 9.0) & $70.7$ (sd: 10.7) \\ 
 Sex Female & $51.6{\%}$ & $47.0{\%}$ \\ 
 Education Secondary & $32.9{\%}$ & $36.4{\%}$ \\ 
 Education College & $37.6{\%}$ & $30.5{\%}$ \\ 
 Education Graduate & $8.3{\%}$ & $5.3{\%}$ \\ 
 Education Missing & $21.3{\%}$ & $27.4{\%}$ \\ 
 Socioeconomic Vulnerability Score (0-1) & $0.33$ (sd: .22) & $0.33$ (sd: .21) \\ 
 Home Life Vulnerability Score (0-1) & $0.45$ (sd: .20) & $0.46$ (sd: .21) \\ 
 Racial Ethnic Vulnerability Score (0-1) & $0.44$ (sd: .22) & $0.42$ (sd: .22) \\ 
 Housing Vulnerability Score (0-1) & $0.51$ (sd: .18) & $0.51$ (sd: .18) \\ 
 Hypertension Diagnostic Code & $63{\%}$ & $66{\%}$ \\ 
 Stroke Diagnostic Code & $4.6{\%}$ & $6.4{\%}$ \\ 
 Cancer Diagnostic Code & $32{\%}$ & $34{\%}$ \\ 
 COPD Diagnostic Code & $3.7{\%}$ & $4.1{\%}$ \\ 
 Overweight Diagnostic Code & $10.4{\%}$ & $4.6{\%}$ \\ 
 Obesity Diagnostic Code & $33.6{\%}$ & $20.5{\%}$ \\ 
 Cardiovascular Disease Diagnostic Code & $18.2{\%}$ & $28.1{\%}$ \\ 
 Visits Before Baseline (mean) & $66.3$ (sd: 94.3) & $47.4$ (sd: 72.1) \\ 
 Visits Year Before Baseline (mean) & $9.86$ (sd: 14.2) & $8.75$ (sd: 14.5) \\ 
 Outpatient Visits Before Baseline (mean) & $58.6$ (sd: 87.5) & $37.5$ (sd: 63.5) \\ 
 Outpatient Visits Year Before Baseline (mean) & $9.12$ (sd: 13.7) & $6.95$ (sd: 12.9) \\ 
 \end{tabular}}
\end{table*}

\FloatBarrier

\section{TTE Details}

\subsection{Outcome Definitions}

ADRD outcomes were defined as the first occurrence of any of the following ICD9, ICD10, internal diagnosis codes, or medications indicating cognitive decline. These sets were developed by consultation with expert clinicians. 
\vspace{5mm}

The ICD9 codes used were the following:
\\
294.8, 290.40, 294.20, 294.21, 290.0, 294.10, 331.83, 331.9, 294.0, 294.9, 290.13, 331.3, 331.0, 331.5, 331.2, 331.82, 290.43, 290.21, 290.10, 780.93, 290, 331, 294, 294.1, 290.41, 290.3, 294.11, 290.20, 290.42, 290.4, 291.2, 290.11, 331.11, 331.89, 290.9, 331.1, 331.19, 331.7, 290.12, 290.0.1, 290.21.1, 290.20.1, 290.40.1, 290.1, 294.80.1, 294.10.1, 290.10.1, 292.82, 290.3.1, 331.2.3, 331.0.3, 290.42.1, 331.81, 290.8, 294.9.1, 290.43.1, 290.2
\vspace{5mm}

The ICD10 codes used were the following:
\\
F03.90, F03.91, F01.50, G31.84, F01.C0, G30.9, G30.1, F02.80, G31.83, G31.89, F01.51, F02.B0, F02.81, F01.518, G30.0, F03.918, F03.A0, G31.9, F10.27, G30.8, I69.311, F02.818, G31.09, F01.A0, F02.A0, F03.911, F03.C0, F01.B0, F02.C11, G31.2, F03.B0, F03.B18, F10.97, I69.911, F03.B11, G31.85, F03.92, F02.C0, I69.811, F02.811, F03.C11, F03.A4, F01.A18, G31.01, G31.81, F01.B11, G31.1, F02.C18, F01.52, F03.9, F03.C18, F02.B2, F01.B4, I69.211
\vspace{5mm}

The other codes used were the following:
\\
LPA99, YHAL6, WLAG8, WHMT3, LPA1009, WLEN6, LPA1404, LPA730, LPA867
\vspace{5mm}

The medications used were the following:
\\
Galantamine (Razadyne, Razadyne ER)
Donepezil (Aricept)
Rivastigmine (Exelon)
Memantine (Namenda)
\vspace{5mm}

Death outcomes were determined by death records within the EHR, supplemented with data from the MA death registry \citep{death}.

\newpage
\subsection{Consort Diagram}

\begin{figure*}[!h]
\floatconts
 {fig:nodes}
 {\caption{Consort diagram showing the number of metformin and sulfonylurea patients excluded at each step in the eligibility criteria filtering. This diagram is for the (B) and (M) variants; for (E), there is a final step applied at the end removing all patients without a PCP indication.}}
 {\includegraphics[width=1\linewidth]{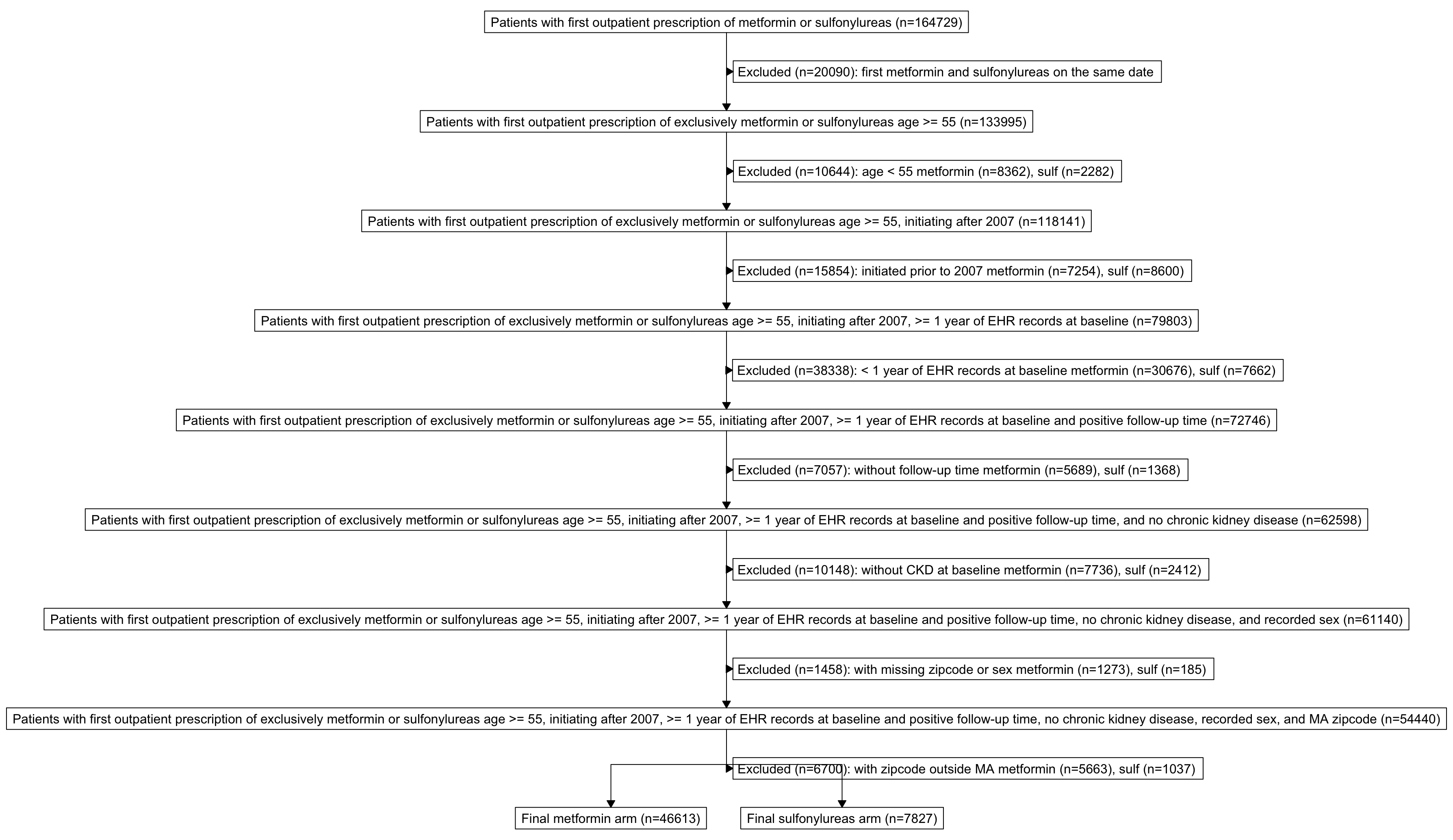}}
\end{figure*}

\FloatBarrier

\subsection{Covariate List}

The covariates used in the propensity model and in the Cox PH model were the following:

\begin{itemize}
 \item Age at baseline
 \item Sex
 \item Hypertension prior to baseline
 \item Stroke prior to baseline
 \item Chronic Obstructive Pulmonary Disease prior to baseline
 \item Overweight diagnosis prior to baseline
 \item Obesity diagnosis prior to baseline
 \item Cardiovascular disease prior to baseline
 \item Cancer (defined with a strict set of ICD codes) prior to baseline
 \item Cancer (defined with a broad set of ICD codes) prior to baseline
 \item Educational attainment level (pre-college, college, graduate, or missing)
 \item Socio-economic vulnerability score (accounts for income, employment, debt, education, etc.)
 \item Home life vulnerability score (accounts for age of family members, size of family, language proficiency in family, other family vulnerabilities, etc.)
 \item Racial/Ethnic vulnerability score (accounts for racial and ethnic minority status)
 \item Housing vulnerability score (accounts for home type, vehicle access, etc.)
 \item BMI classification ($0-<25$, $25-<30$, $30+$)
\end{itemize}

These covariates were used as main effects.

The social vulnerability index score values were determined by the mean SVI values from all the census tracts within the patient's zip code, using data from the Agency for Toxic Substances and Disease Registry \citep{svi}. Scores range from 0 to 1, where higher scores indicate more vulnerability.

\newpage
\section{Full Forest Plots from the Cox PH Models}

Forest plots depicting the estimated hazard ratios from the Cox proportional hazards models for all three variants. The first line shows the estimated treatment effect; following lines show estimated hazard ratios for the covariates in the model.

\begin{figure*}[!h]
\floatconts
 {fig:nodes}
 {\caption{Forest Plot of Cox PH Model (B)}}
 {\includegraphics[width=1\linewidth]{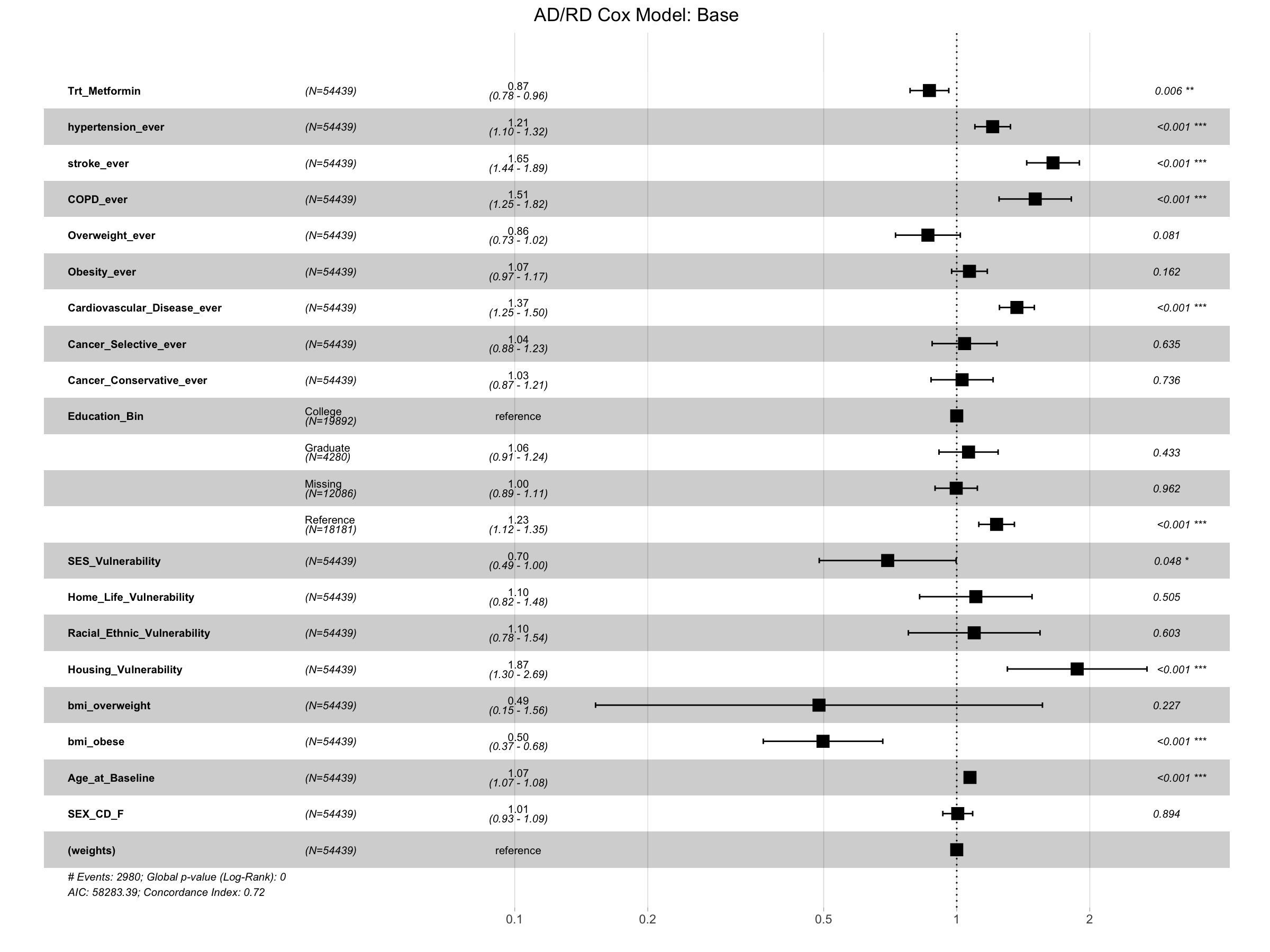}}
\end{figure*}

\begin{figure*}[!h]
\floatconts
 {fig:nodes}
 {\caption{Forest Plot of Cox PH Model (M)}}
 {\includegraphics[width=1\linewidth]{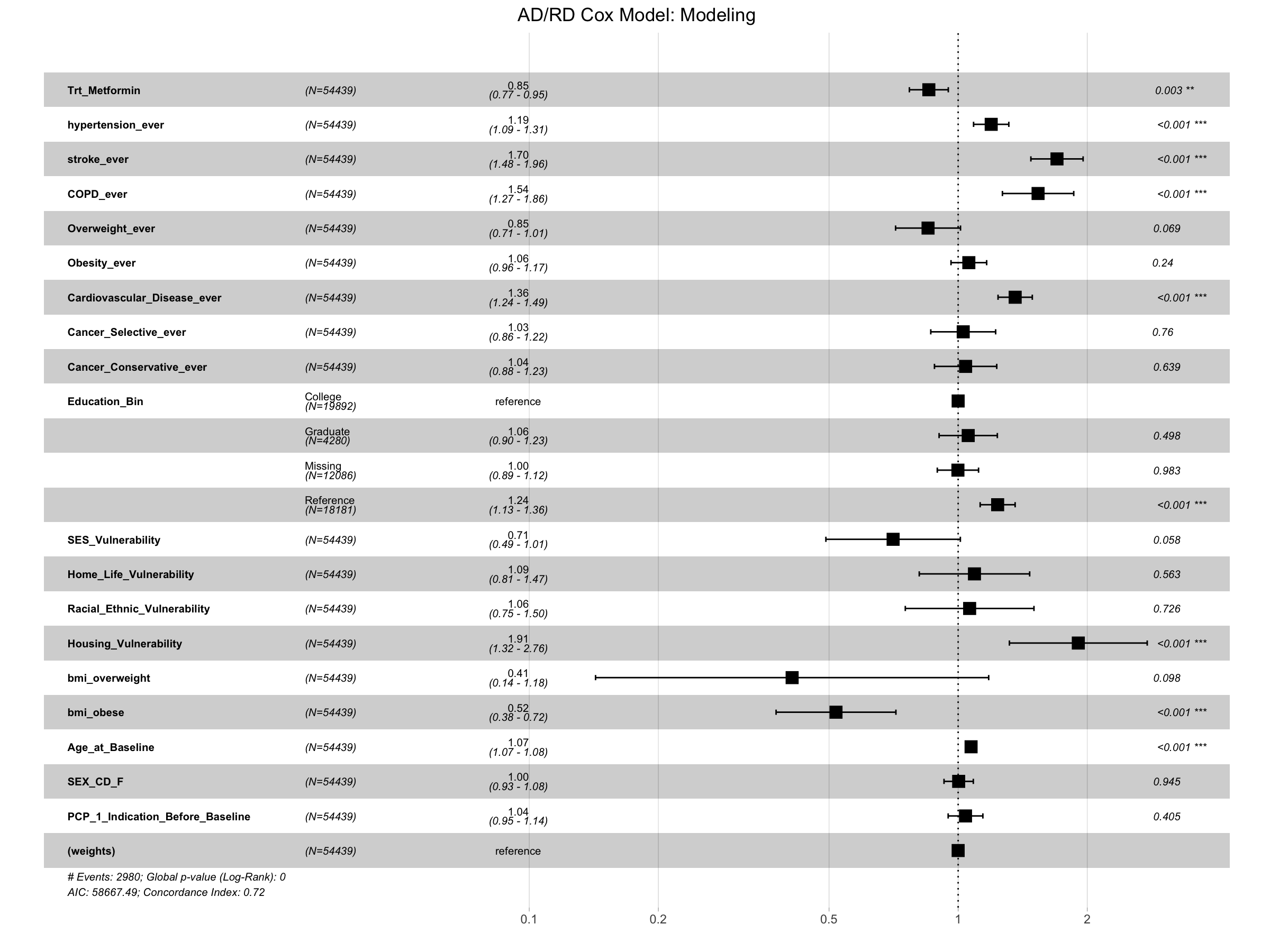}}
\end{figure*}

\begin{figure*}[!h]
\floatconts
 {fig:nodes}
 {\caption{Forest Plot of Cox PH Model (E)}}
 {\includegraphics[width=1\linewidth]{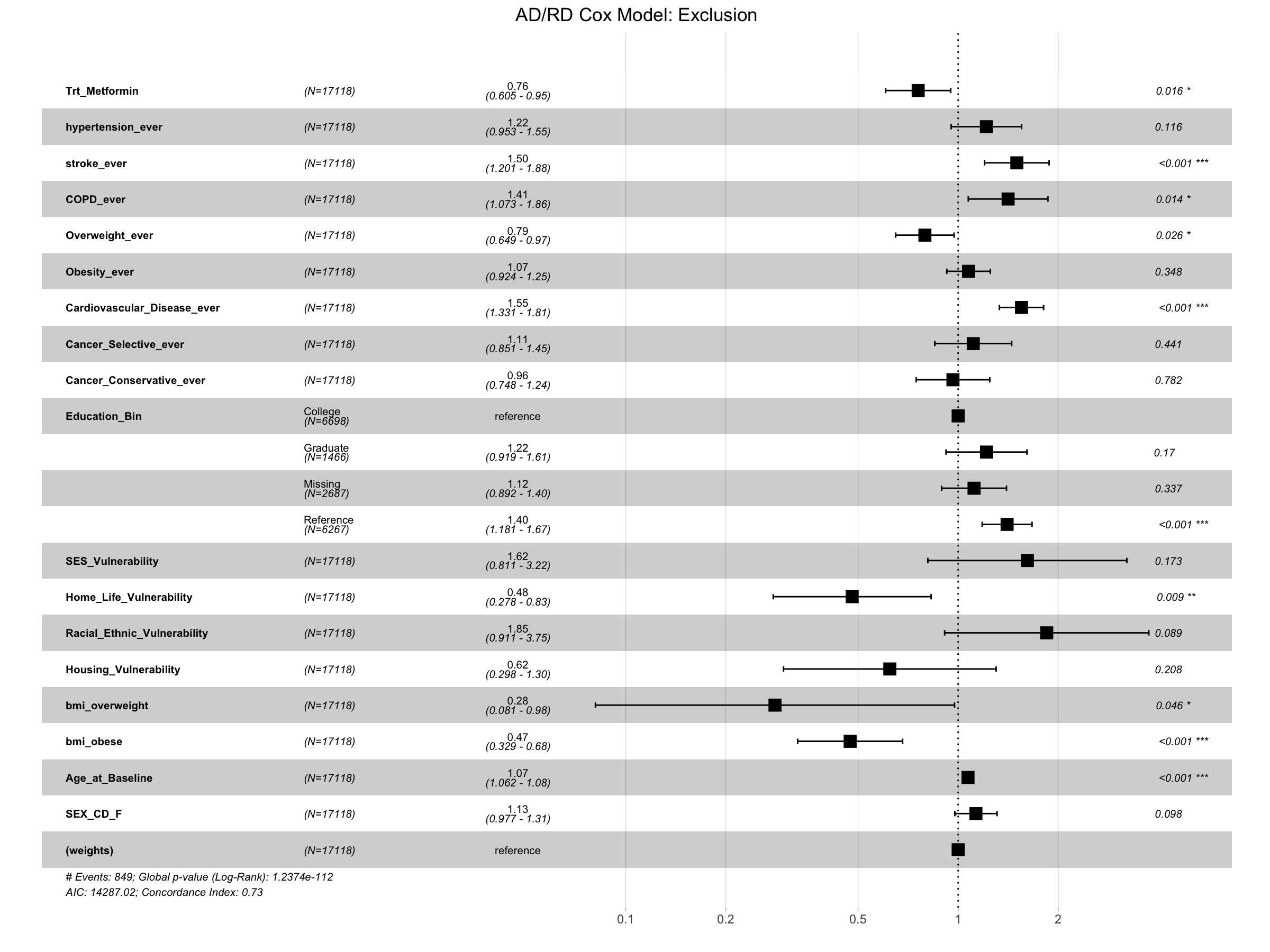}}
\end{figure*}

\FloatBarrier

\section{Full-size Individual CIFs}

\begin{figure*}[!h]
\floatconts
 {fig:nodes}
 {\caption{CIFs accounting for competing risks (B)}}
 {\includegraphics[width=1\linewidth]{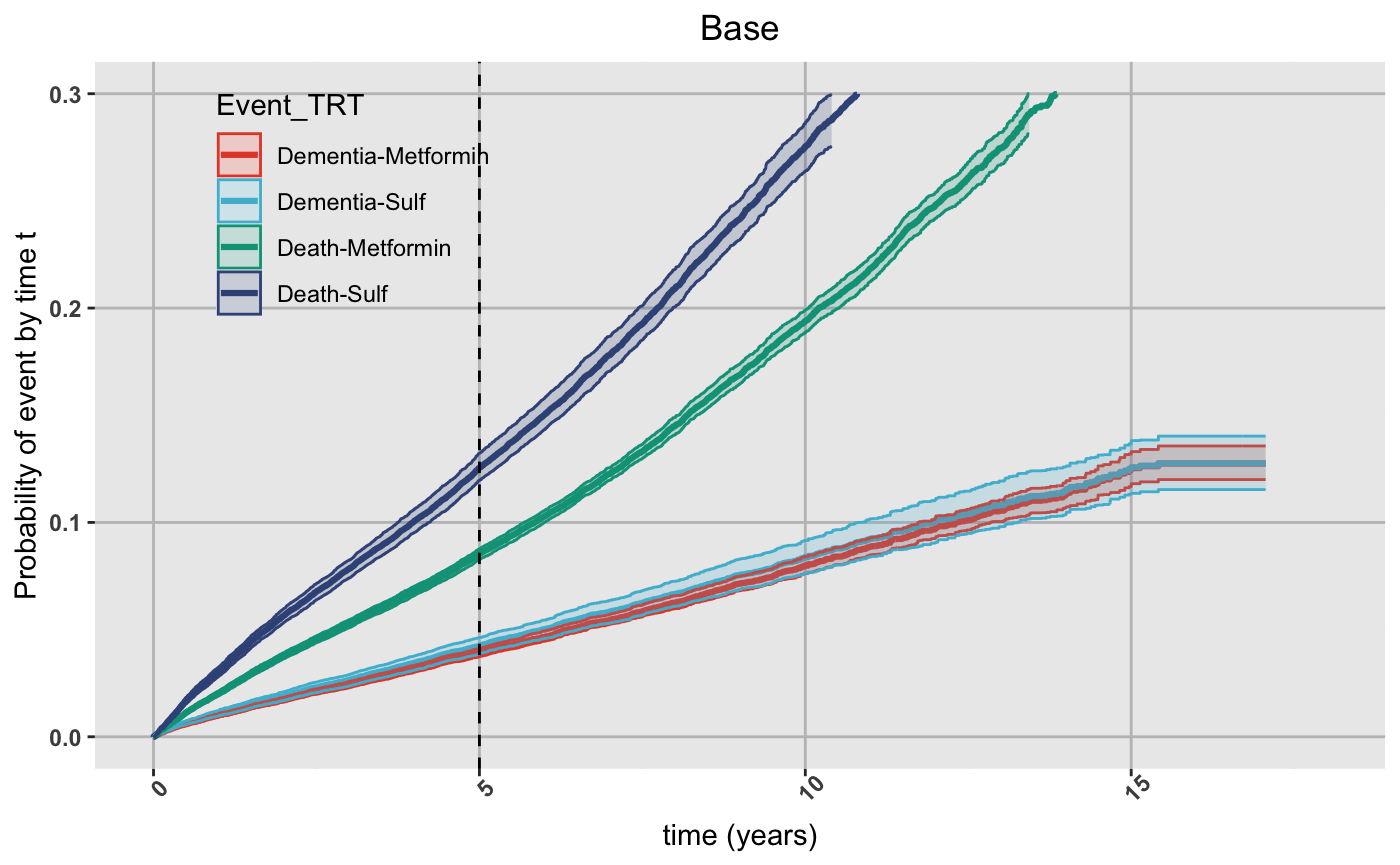}}
\end{figure*}

\begin{figure*}[!h]
\floatconts
 {fig:nodes}
 {\caption{CIFs accounting for competing risks (M)}}
 {\includegraphics[width=1\linewidth]{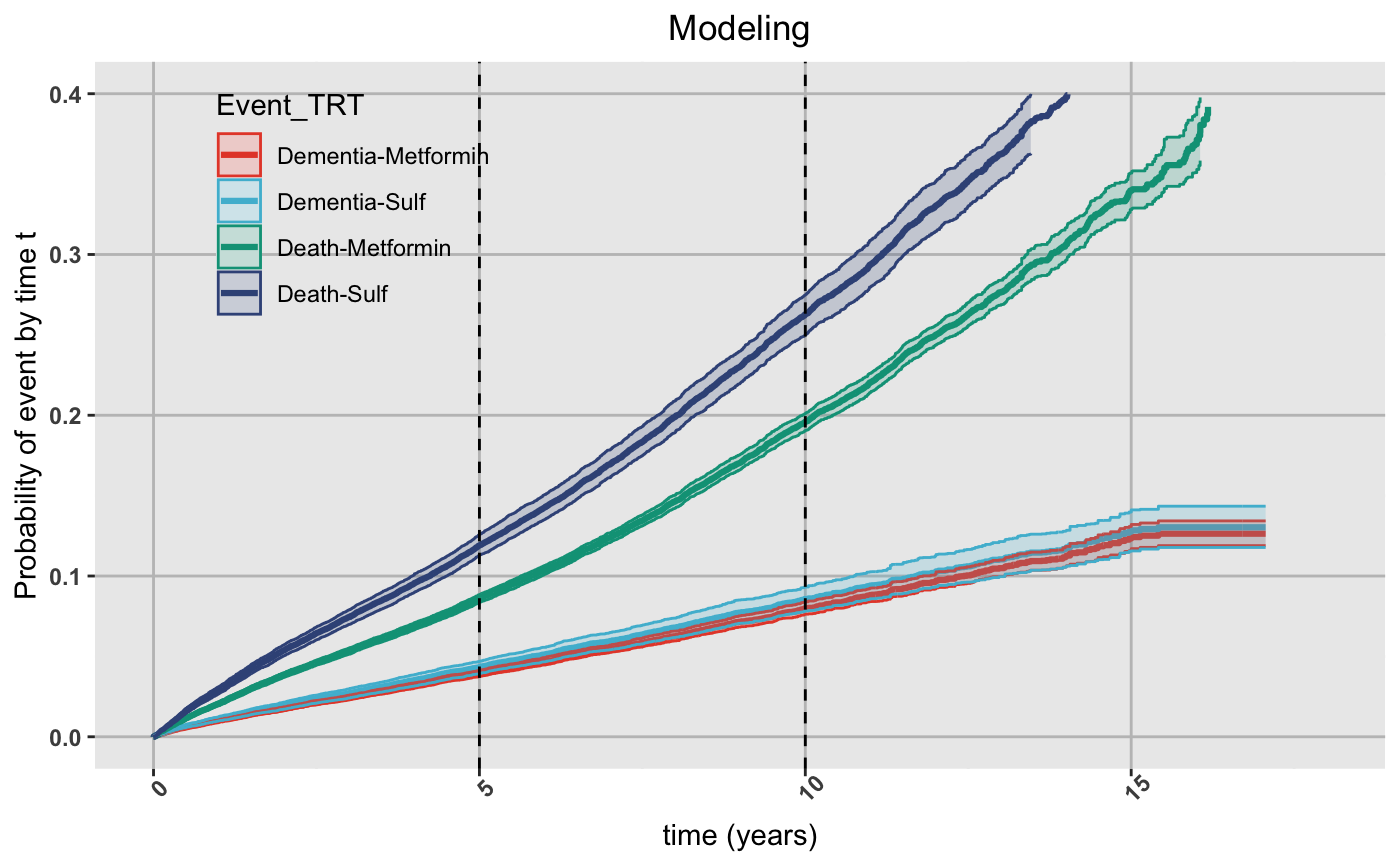}}
\end{figure*}

\begin{figure*}[!h]
\floatconts
 {fig:nodes}
 {\caption{CIFs accounting for competing risks (E)}}
 {\includegraphics[width=1\linewidth]{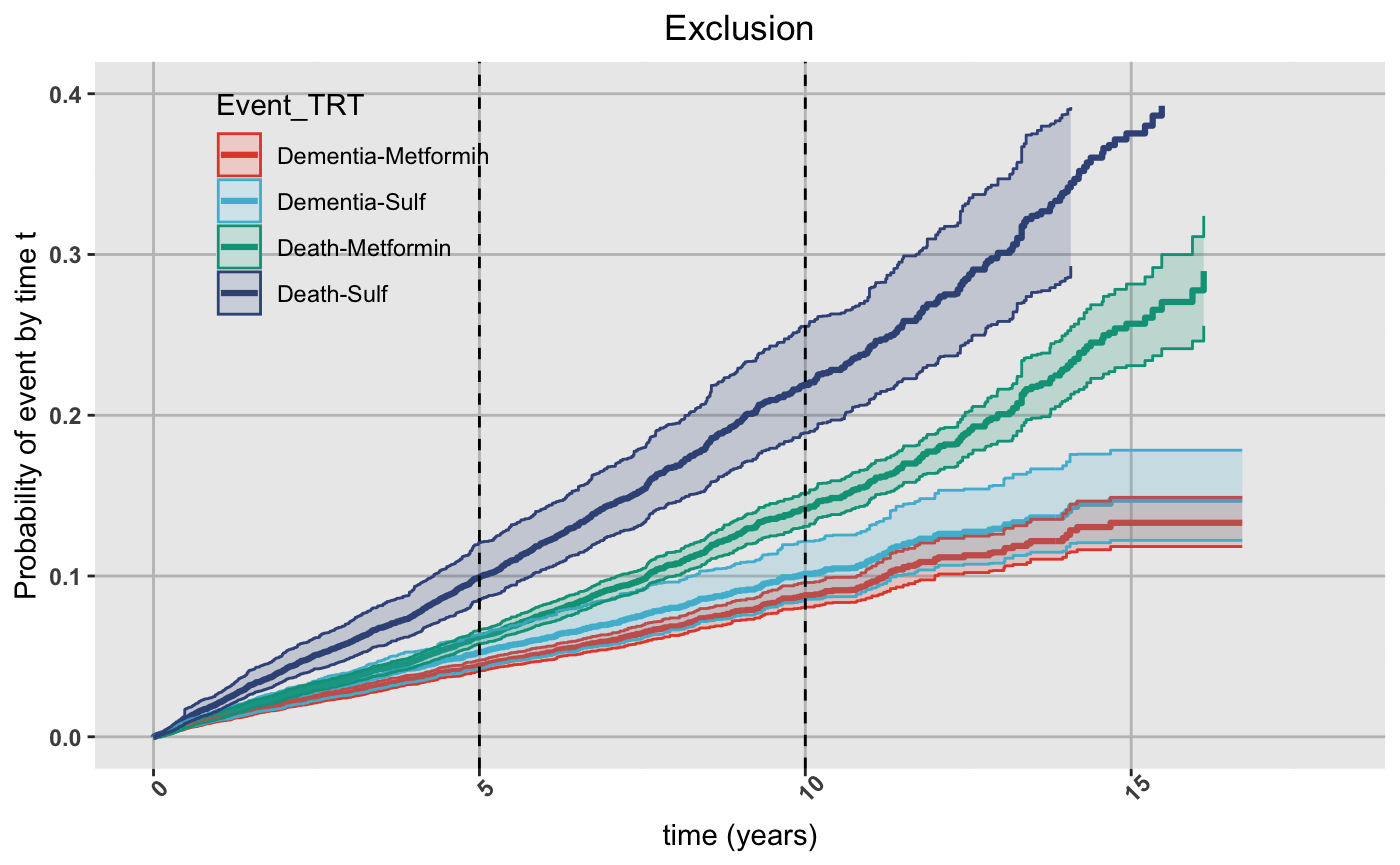}}
\end{figure*}

\FloatBarrier

\section{Risk Difference Plots}

\begin{figure*}[htbp]
\floatconts
 {fig:nodes}
 {\caption{Risk Differences over time for ADRD and D{\O}D (B)}}
 {\includegraphics[width=.8\linewidth]{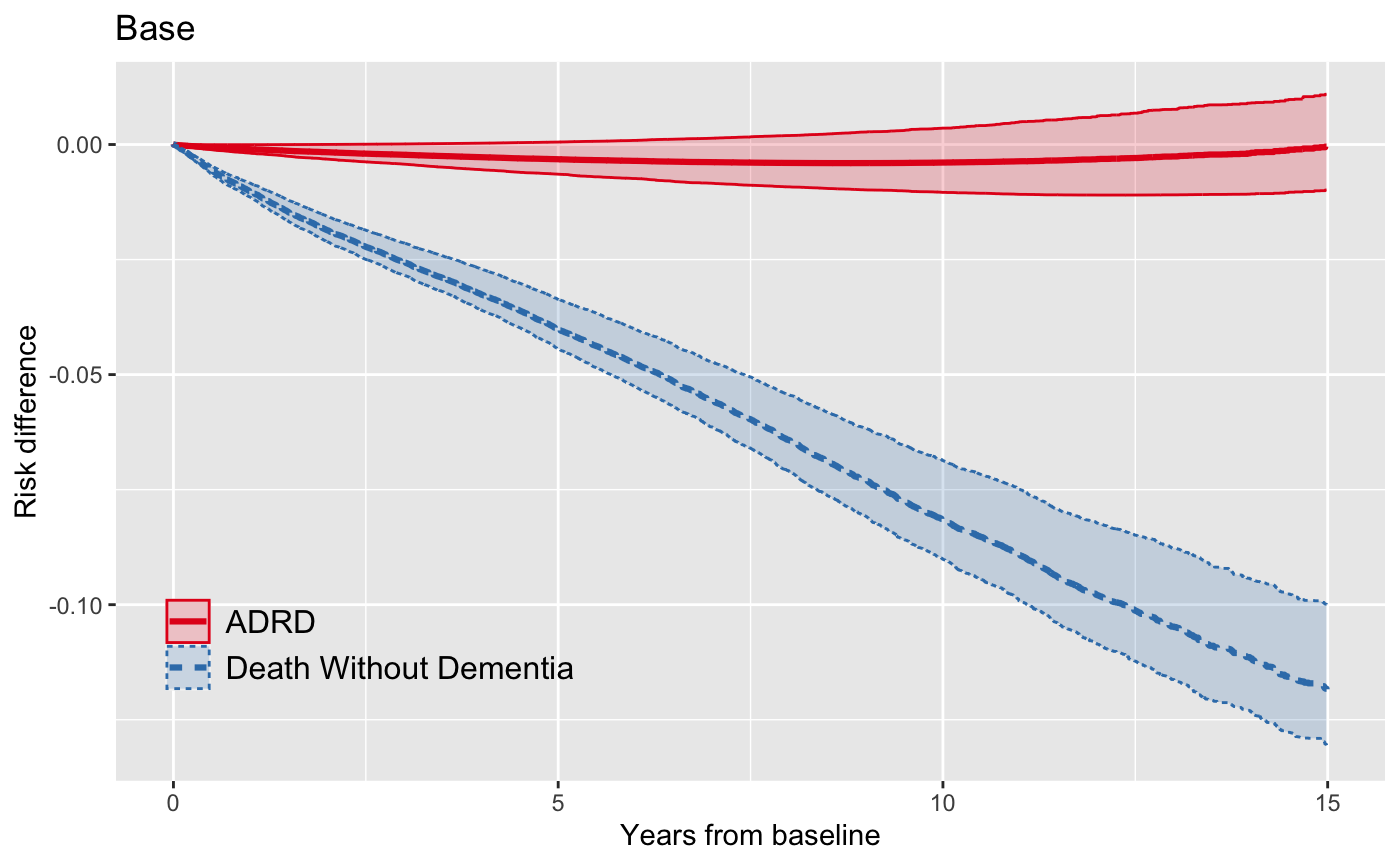}}
\end{figure*}

\begin{figure*}[htbp]
\floatconts
 {fig:nodes}
 {\caption{Risk Differences over time for ADRD and D{\O}D (M)}}
 {\includegraphics[width=.8\linewidth]{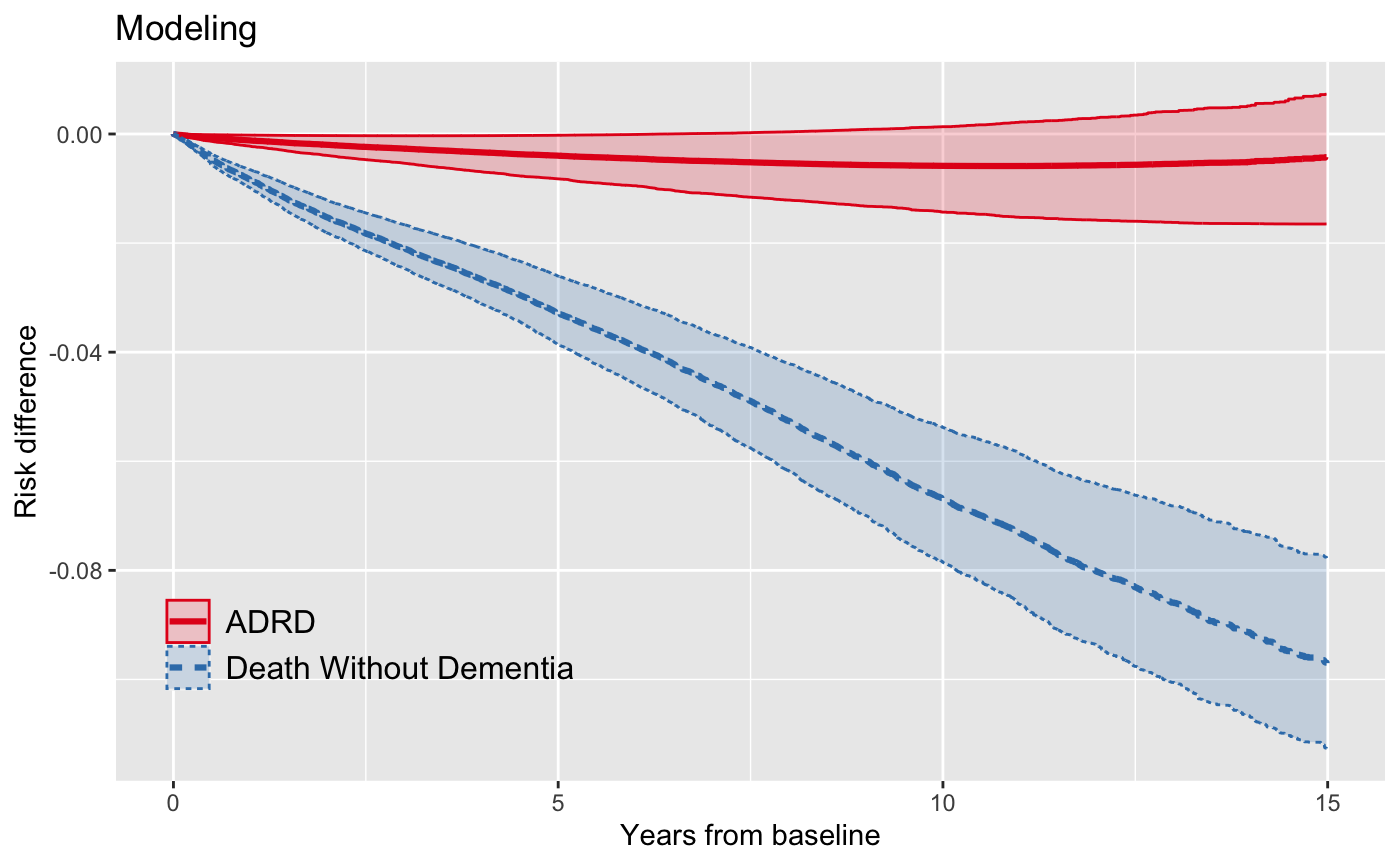}}
\end{figure*}

\begin{figure*}[htbp]
\floatconts
 {fig:nodes}
 {\caption{Risk Differences over time for ADRD and D{\O}D (E)}}
 {\includegraphics[width=.8\linewidth]{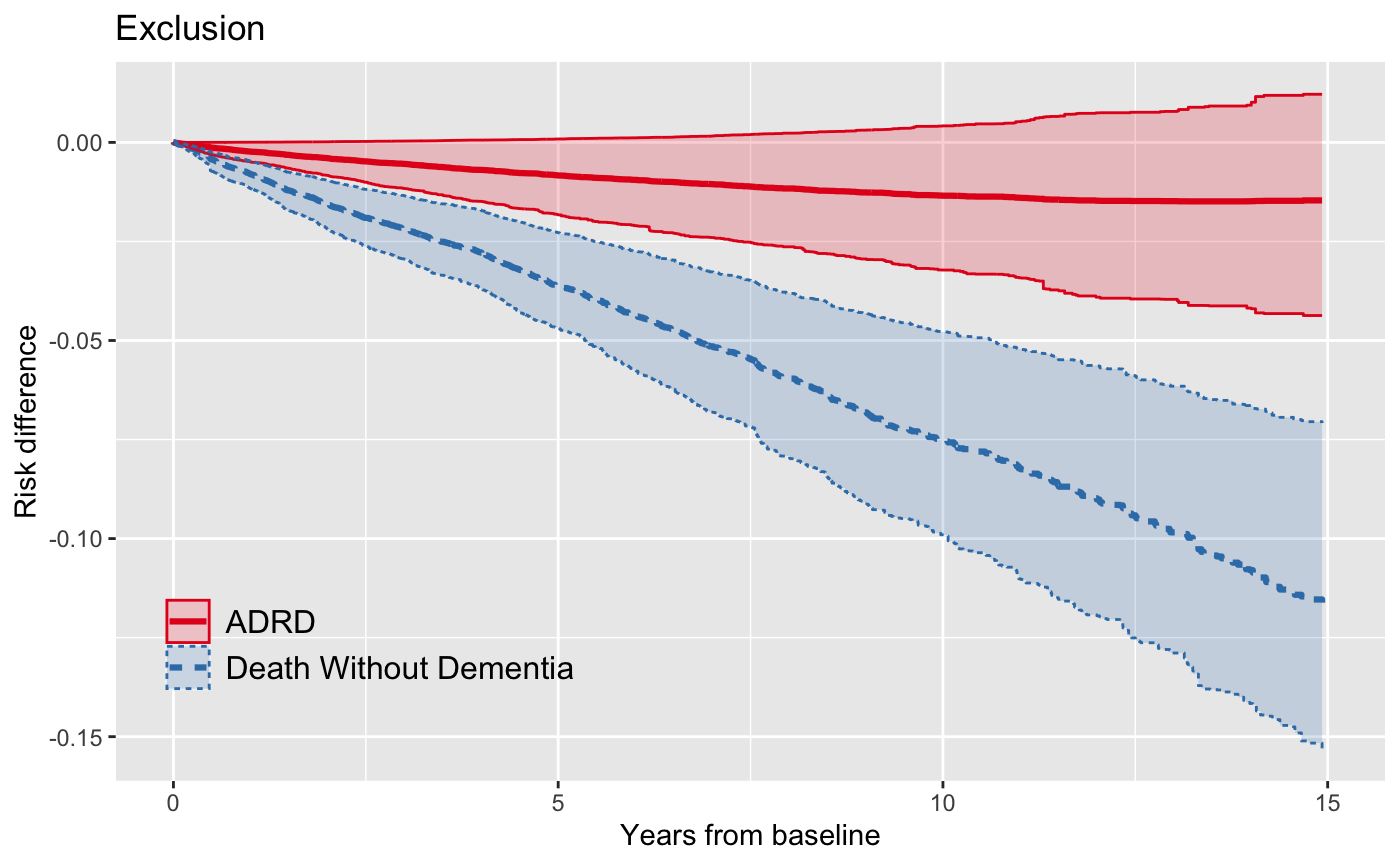}}
\end{figure*}

\section{Propensity Score Distributions}

Plots showing the distribution of propensity scores grouped by treatment arm. When a score is closer to $1$, the model is more confident that the patient is in the metformin arm. The distribution for metformin patients (red) is closer to the right than that of the sulfonylurea patients (blue), but there is significant overlap, suggesting that the two arms are composed of comparable populations.  To calculate weights from the propensity scores, we use stabilized average treatment effect weighting (as did the original metformin vs. sulfonylureas TTE).

\begin{figure*}[!h]
\floatconts
 {fig:nodes}
 {\caption{Propensity Distributions (B)}}
 {\includegraphics[width=1\linewidth]{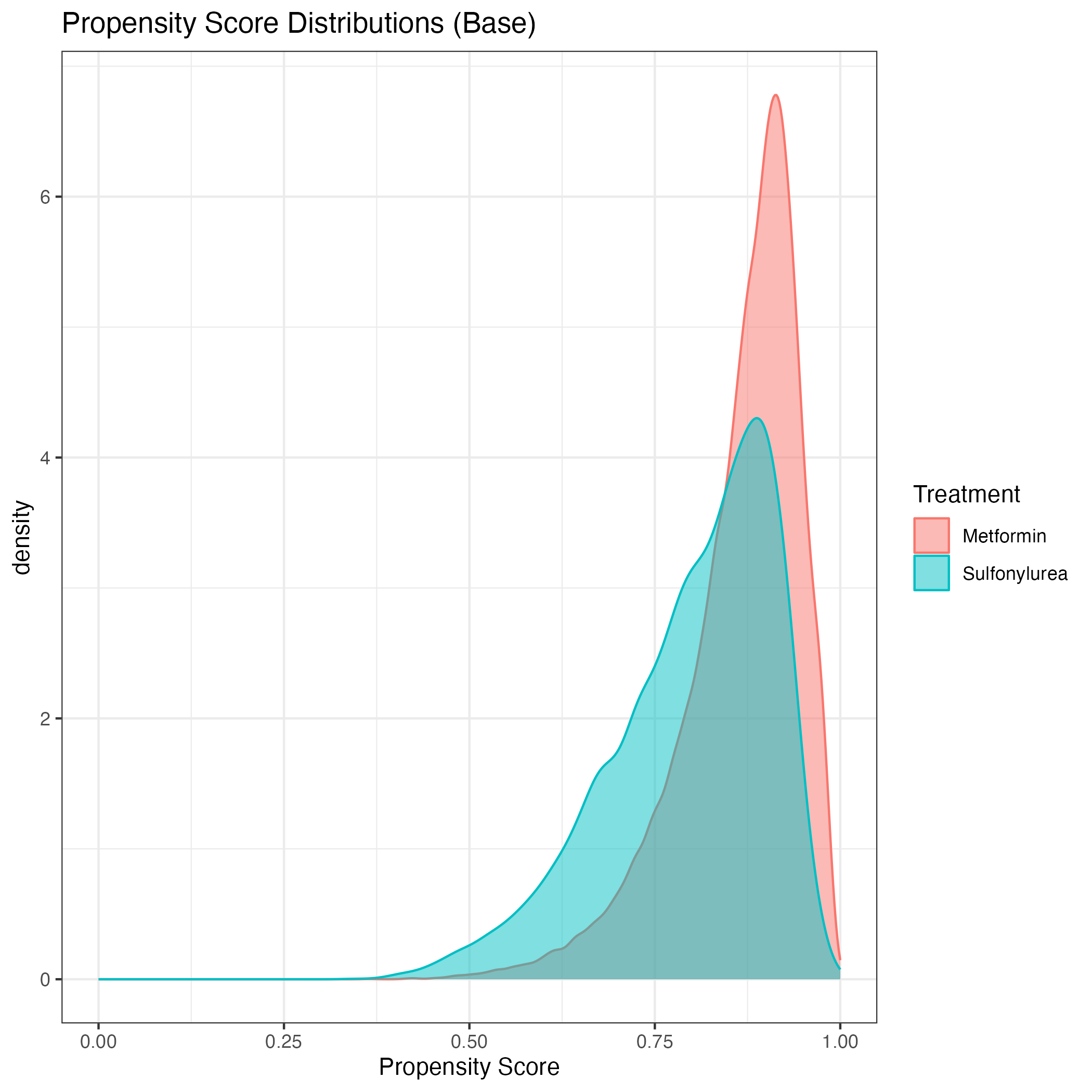}}
\end{figure*}

\begin{figure*}[!h]
\floatconts
 {fig:nodes}
 {\caption{Propensity Distributions (M)}}
 {\includegraphics[width=1\linewidth]{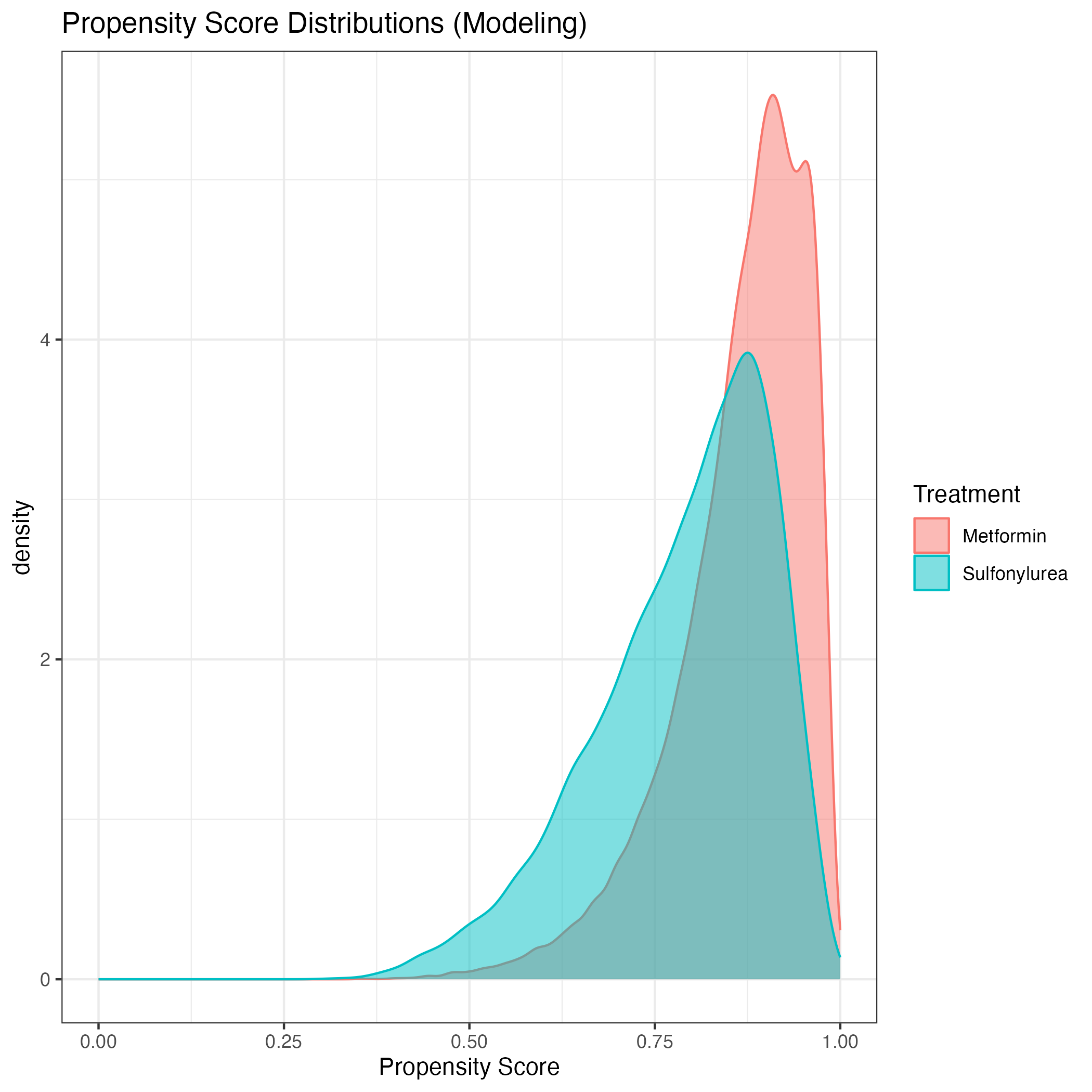}}
\end{figure*}

\begin{figure*}[!h]
\floatconts
 {fig:nodes}
 {\caption{Propensity Distributions (E)}}
 {\includegraphics[width=1\linewidth]{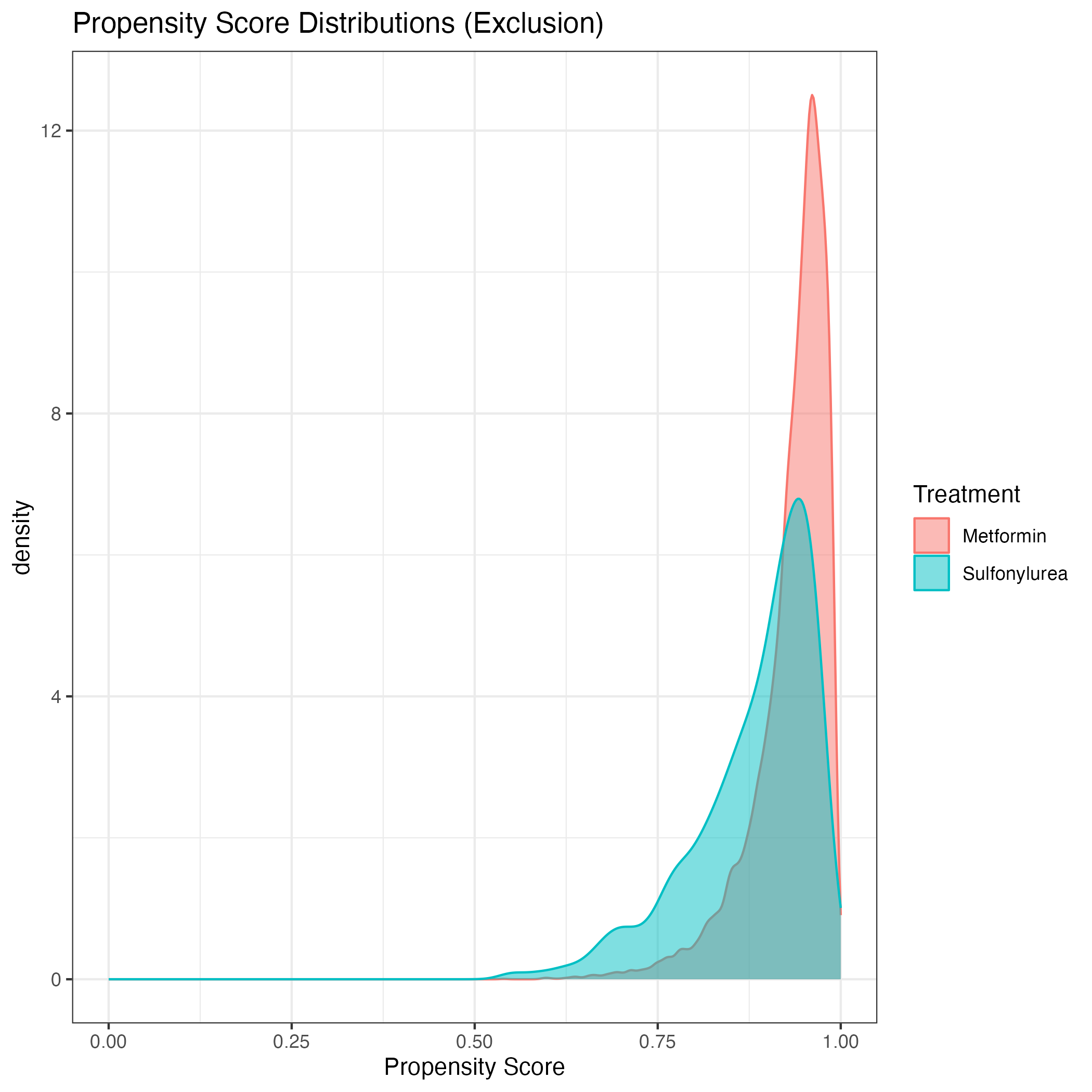}}
\end{figure*}

\FloatBarrier

\section{Age-specific ADRD Incidence per Person-year}

The age-specific ADRD incidence per person year. Black trend line is constructed by interpolating between data points extracted from a figure reported in a study of over 8 million patients using medicare claims data \citep{olfson-21}. For the ADRD incidence rates in our cohort, we use the same ICD9 code set as the reference paper with the addition of the analogous ICD10 codes.

\begin{figure*}[!h]
\floatconts
 {fig:nodes}
 {\caption{Age-specific ADRD Incidence per Person-year}}
 {\includegraphics[width=1\linewidth]{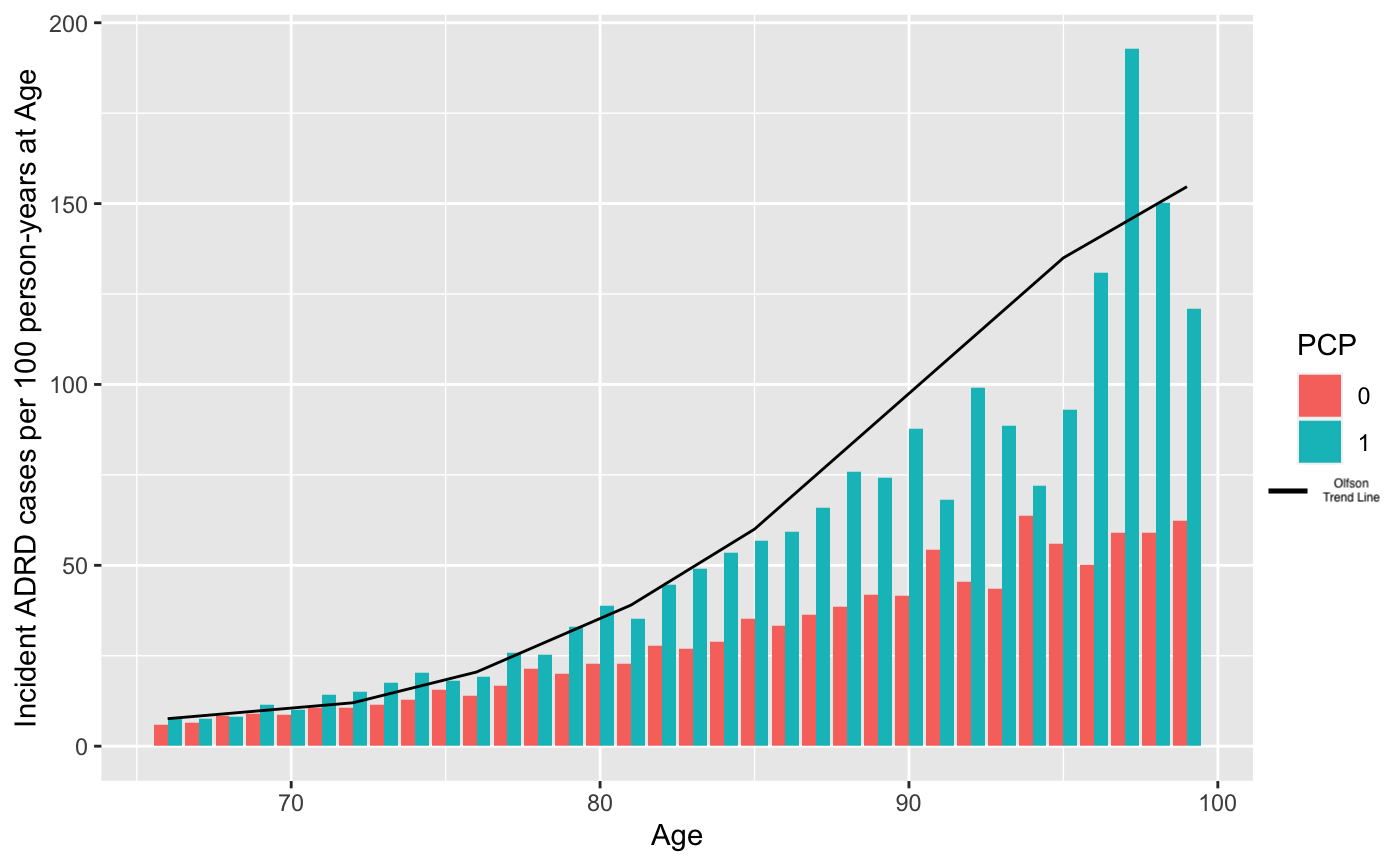}}
\end{figure*}

\FloatBarrier

\newpage
\section{Age-specific mortality Incidence per Person-year}

The age-specific mortality incidence per person year, stratified by whether a patient has an internal PCP utilization indication prior to study initiation, only post study initiation, or never. Black trend line illustrates the incidences reported in the Massachusetts life tables \citep{arias-22}.

\begin{figure*}[!h]
\floatconts
 {fig:nodes}
 {\caption{Age-specific mortality Incidence per Person-year}}
 {\includegraphics[width=1\linewidth]{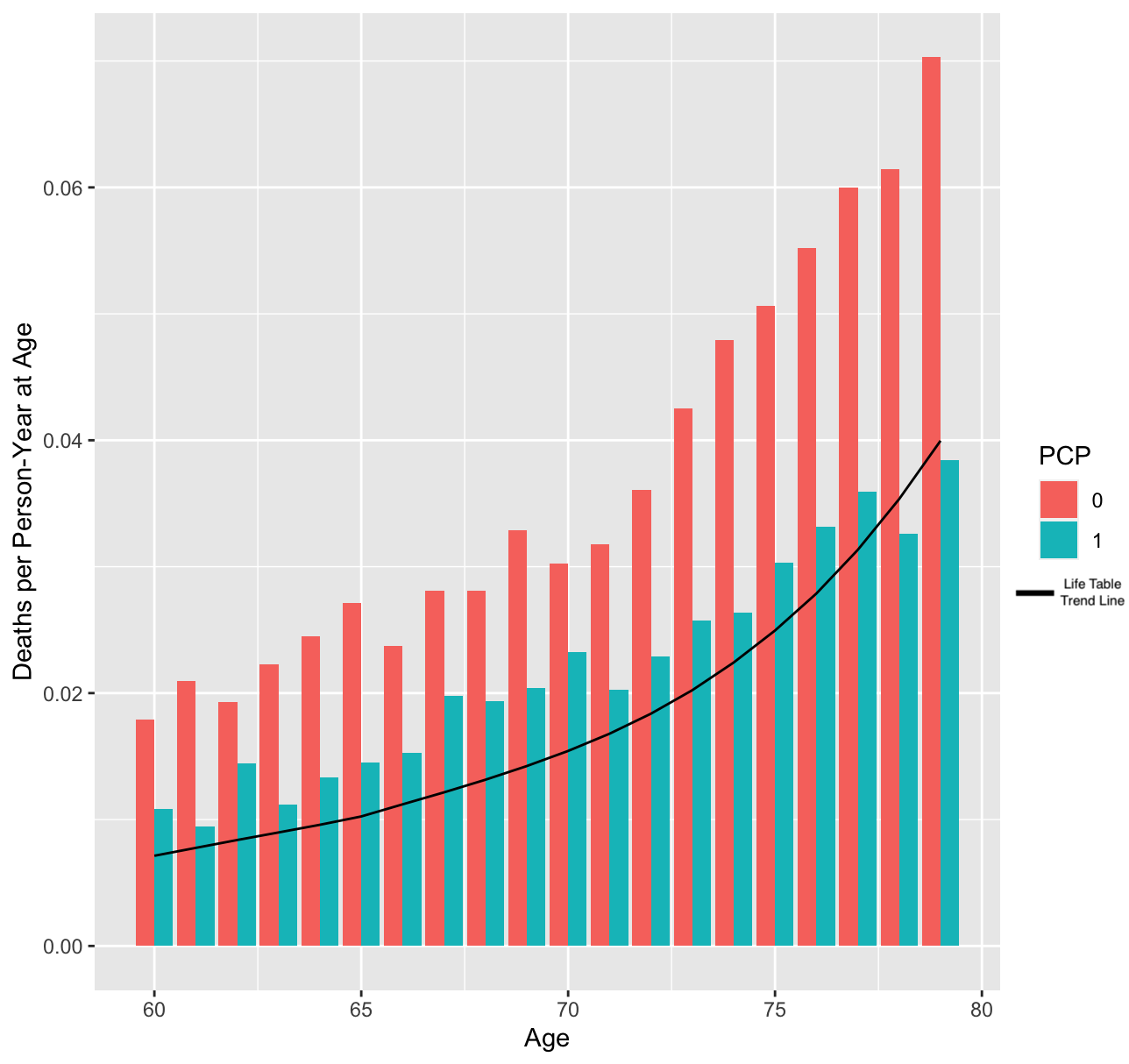}}
\end{figure*}

\FloatBarrier

\end{document}